\documentclass{article}
\usepackage{spconf,amsmath,graphicx,multirow,hhline,booktabs}
\usepackage[hidelinks]{hyperref}
\usepackage{enumitem}
\setlist{nosep,leftmargin=14pt}

\usepackage[capitalize]{cleveref}
\usepackage{xcolor}

\title{Benchmarking Hierarchical Image Pyramid Transformer for the classification of colon biopsies and polyps in histopathology images}

\name{\parbox{\linewidth}{\centering
Nohemí S. León Contreras$^{1}$, Clément Grisi$^{1}$, Witali Aswolinskiy$^{1}$, Simona Vatrano$^{2}$, Filippo Fraggetta$^{2}$, Iris Nagtegaal$^{1}$, Marina D'Amato$^{1*}$, Francesco Ciompi$^{1*}$\thanks{*Authors equally contributed in supervising this work}
}}

\address{
$^{1}$ Department of Pathology, Radboud University Medical Center, Nijmegen, The Netherlands \\
$^{2}$ Pathology Unit, Gravina Hospital, Caltagirone, Italy
}

\begin{document}
\maketitle
\begin{abstract}
Training neural networks with high-quality pixel-level annotation in histopathology whole-slide images (WSI) is an expensive process due to gigapixel resolution of WSIs.
However, recent advances in self-supervised learning have shown that highly descriptive image representations can be learned without the need for annotations. 
We investigate the application of the recent Hierarchical Image Pyramid Transformer (HIPT) model for the specific task of classification of colorectal biopsies and polyps.
After evaluating the effectiveness of TCGA-learned features in the original HIPT model, we incorporate colon biopsy image information into HIPT's pretraining using two distinct strategies: (1) fine-tuning HIPT from the existing TCGA weights and (2) pretraining HIPT from random weight initialization. We compare the performance of these pretraining regimes on two colorectal biopsy classification tasks: binary and multiclass classification.
\end{abstract}
\begin{keywords}
Computational pathology, colon biopsies, self-supervision, vision transformers.
\end{keywords}

\vspace{-0.3cm}
\section{Introduction}\label{sec:intro}
\vspace{-0.1cm}
With the digitization of histopathology images, recent advancements in whole-slide image (WSI) analysis have largely focused on deep learning (DL) as the core technology for decision support tools in clinical diagnostics, aiming at easing the workload of clinicians \cite{abels2019computational, laurinavicius2012digital}.
However, the sheer gigapixel dimensions of WSIs make end-to-end supervised machine learning approaches impractical\cite{vanderLaak2021DeepLearning}, with applications commonly resorting to subdividing WSIs into smaller patches for analysis.
Furthermore, obtaining adequate, high-quality pixel-level annotated datasets in histopathology represents a resource-intensive and time-consuming process.
To address this challenge, weakly-supervised algorithms have emerged to leverage only slide-level labels, often derived from clinically available data such as pathology reports or patient clinical histories \cite{lu2021CLAM}.
Some weak supervision techniques leverage pretrained encoders to generate embeddings from tissue patches \cite{chen2022self, ilse2018attention}.
Historically based on models pretrained on the ImageNet dataset via full supervision and used as feature extractors, recent developments in weakly supervised learning have seen the advent of self-supervised learning methods, allowing to reduce the gap in domain shift by using domain-specific unlabeled data for pretraining models, which are then capable of yielding meaningful high-level feature representations of the data \cite{chen2022self}. 

\begin{figure}[t]
    \centering
    \includegraphics[width=1.0 \linewidth]{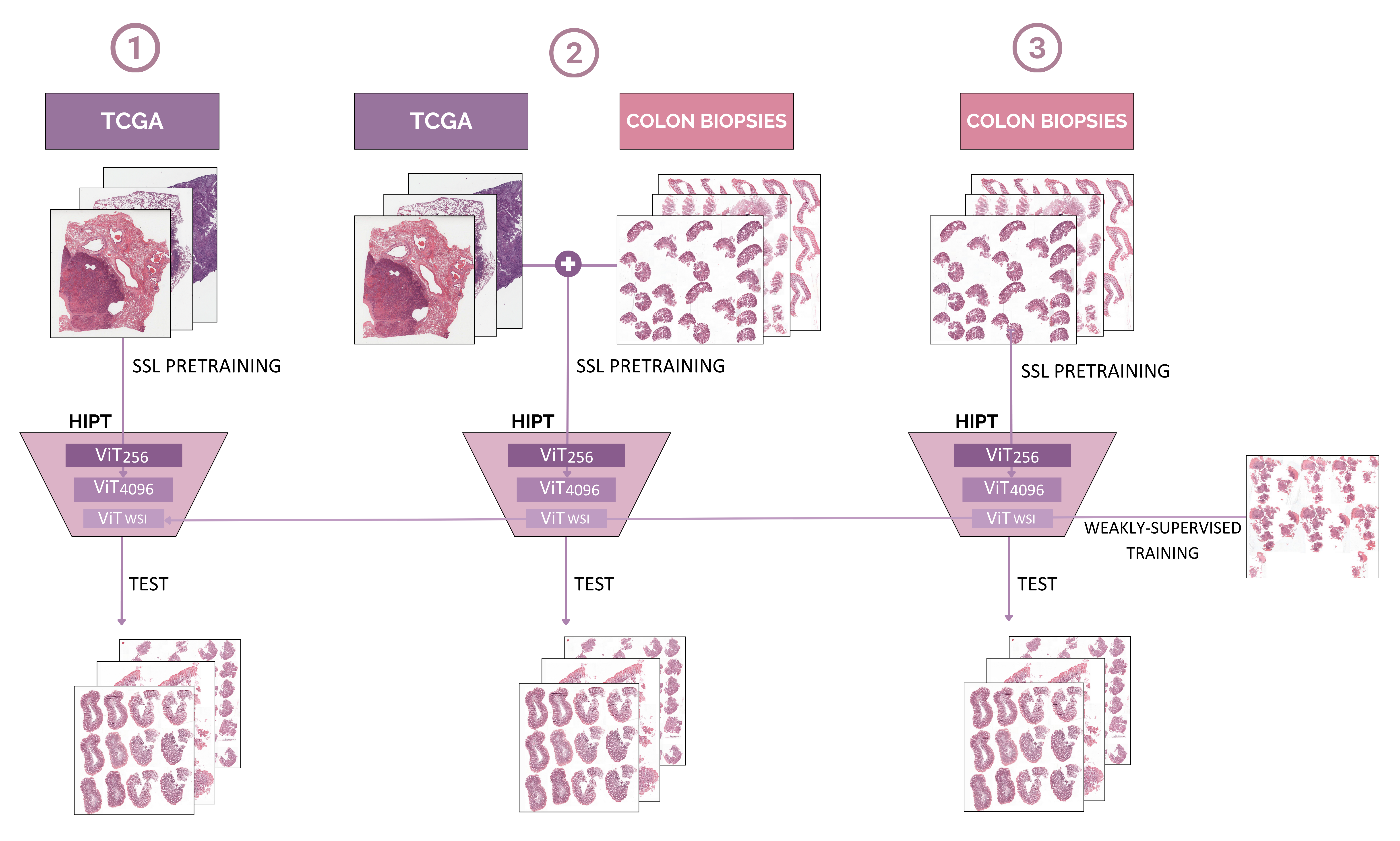} 
  \caption{Schematic overview of the scenarios we investigate in this paper. 1) we take an ``off-the-shelf'' HIPT pretrained on TCGA via self-supervised learning (SSL) as described in \cite{chen2022scaling}; 2) we initialize HIPT with TCGA-pretrained weights and further train it via SSL using an application-specific dataset of colon biopsies; 3) we pretrain HIPT from scratch via SSL solely using colon biopsies data. In all scenarios, we train the final level of HIPT (i.e., a ViT classifier) using colon biopsy whole-slide images.}
  \label{fig:overview}
\end{figure}

In recent years, Vision Transformers (ViT) are replacing convolutional neural networks as the adopted architecture to pretrain with self-supervised learning (SSL).
A recent incarnation of such a trend is the recently introduced Hierarchical Image Pyramid Transformer (HIPT) proposed by \cite{chen2022scaling} to take advantage of the intrinsic pyramidal structure of WSI data and adopts multiple stages of ViTs pretrained via SSL on histopathology images. In particular, HIPT aggregates visual tokens at incrementing resolution via two levels of DINO-pretrained ViTs and a weakly supervised ViT to learn high-resolution image representation. 
The multiple levels of the original HIPT model were pretrained on over 10,000 tissue resections from multiple cancer types from the public The Cancer Genome Atlas (TCGA), while the last level was trained as a classifier with full supervision to address the task at hand.

In this work, we investigate how effectively HIPT leverages knowledge gained from a diverse set of cancer types to address the application-specific task of colorectal biopsy classification.
For this, we explore three different learning scenarios (see Figure \ref{fig:overview}), where 1) we consider the original HIPT model, pretrained solely using TCGA resections; 2) we integrate task-specific information by finetuning HIPT's hierarchical self-supervised pretraining on data from colorectal biopsies, to investigate potential improvements gained by merging knowledge from diverse cancer types (TCGA) with data from the specific task; 3) we assess HIPT's performance when pretrained from scratch on colorectal biopsy data using SSL, aiming to discern the impact of TCGA data on improving generalizability.
In all cases, we train the last level of HIPT to encompass both binary and multiclass classification scenarios.  

\vspace{-0.3cm}
\section{Method}
\vspace{-0.1cm}
In this section, we introduce (1) the data used in this paper, (2) the hierarchical self-supervised pretraining approach, and (3) the slide-level weak supervision tasks we propose to evaluate the learned representations.
\vspace{-0.3cm}
\subsection{Material}
\vspace{-0.1cm}
We used 8,868 H\&E-stained colorectal biopsy WSIs cut from 6,563 paraffin blocks of 3,601 patients from Radboud University Medical Center (RUMC), Nijmegen (The Netherlands), along with their respective pathology report. Based on reports, for each case, we extracted the following labels: normal, hyperplastic polyps, low-grade dysplasia (LGD), high-grade dysplasia (HGD), and cancer. 
To remove excessive background and since the diagnostic reports refer to the tissue present in the whole paraffin blocks, we combined the multiple slides per patient derived from the same block into a single ''macro slide'' using an in-house developed open-source tool\footnote{https://github.com/DIAGNijmegen/pathology-whole-slide-packer}.
Furthermore, a set of 76 colorectal slides with their associated reports from Azienda Ospedaliera Cannizaro and Gravina Hospital Caltagirone ASP, Catania, Italy was used as part of the test set in a classification downstream task (slide-level weak supervision) to test the distribution shift robustness of the features learned through self-supervised pretraining.

\begin{figure}[!th]
    \centering
    \includegraphics[width=1.0 \linewidth]{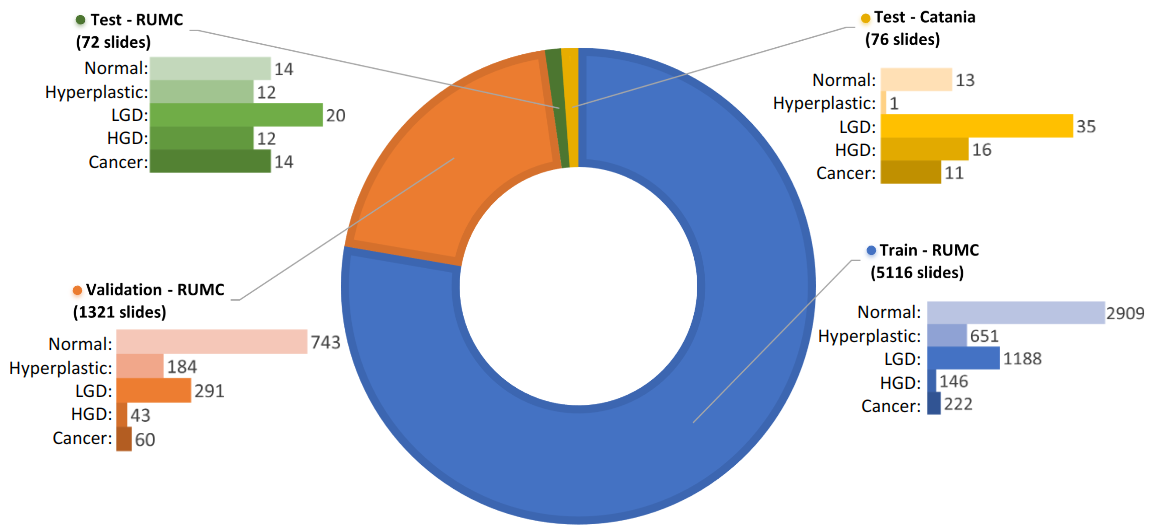} 
  \caption{Data split done for slide-level weak supervision; LGD/HGD: low-grade/high-grade dysplasia.}
  \label{fig:class-split}
\end{figure}

\vspace{-0.3cm}
\subsection{Hierarchical Self-Supervised Pretraining}
\vspace{-0.1cm}
The HIPT architecture adopts a two-level approach, using DINO-based knowledge distillation with ViT to learn high-resolution hierarchical image representation\cite{caron2021DINOemerging,chen2022scaling}.

\textbf{First Level ViT ($ViT_{256}$)}: This level processes $[256\times256]$ pixel image patches by aggregating information from $[16\times16]$ non-overlapping pixel tokens \cite{dosovitskiy2020VIT}.  To model local clusters of cell-to-cell interactions, we extracted tissue containing patches at 20$\times$ magnification \cite{chen2022scaling}. We pretrained $ViT_{256}$ using DINO as proposed by \cite{caron2021DINOemerging}. The resulting class ([CLS]) token from  $ViT_{256}$ is used as the feature representation of the $[256\times256]$ patch.

\textbf{Second Level ViT ($ViT_{4096}$)}: Building upon the representations learned at the first level, $ViT_{4096}$ aggregates non-overlapping $[256\times256]$ patches within $[4096\times4096]$ image regions. This allows $ViT_{4096}$ to characterize macro-scale interactions between cell clusters and their organization in tissue.
The DINO pretraining of $ViT_{4096}$ differs from that of $ViT_{256}$ by embedding the corresponding $[CLS]_{256}$ (subindex denotes the image size represented by the [CLS] token) token feature vectors from the $[4096\times4096]$ image region rather than the raw image patches.

\vspace{-0.3cm}
\subsection{Slide-Level Weak Supervision}
\vspace{-0.1cm}
The final ViT in HIPT, $ViT_{WSI}$, aggregates the region-level representations obtained from $ViT_{4096}$.In this work, $ViT_{WSI}$ undergoes training via binary and four-class classification tasks based on $[CLS]_{4096}$ tokens.

For the binary classification, normal and hyperplastic colorectal WSIs were grouped into a 'Benign' class, whereas LGD, HGD and colorectal cancer WSIs were grouped into an 'Abnormal' class.
For the four-class classification, normal and hyperplastic colorectal WSI were grouped into a 'Benign' class and the other three classes were LGD, HGD, and colorectal cancer.
The macro slides from RUMC were split into train (5116), validation (1321), and test (72) sets. The WSIs from Catania (76) were used exclusively for testing, hence making the test set a total of 148 cases. In \cref{fig:class-split} the distribution of colorectal slides from each class and each center per split is shown. 

\vspace{-0.3cm}
\section{Experiments}
\vspace{-0.1cm}
\subsection{Hierarchical Self-Supervised Pretraining}
\vspace{-0.1cm}
We systematically evaluated several pretraining strategies for both $ViT_{256}$ and $ViT_{4096}$. For the sake of clarity, in our notation the dataset(s) used for pretraining are prefixed before the ViT notation, with TCGA $ViT_{256}$ and TCGA $ViT_{4096}$ representing the original models pretrained only on TCGA data as presented in \cite{chen2022scaling}. 
For the patch-level $ViT_{256}$, we initiated the pretraining process by extracting tissue patches of size $[256\times256]$ from the RUMC dataset at 20X magnification following the methodology proposed in \cite{lu2021CLAM}, resulting in a training dataset of $\sim88M$ patches. 
We then compared two pretraining strategies:
\begin{enumerate}
\item \texttt{TCGA+RUMC $ViT_{256}$} $\to$ finetuning $ViT_{256}$ from the available $TCGA\textnormal{ }ViT_{256}$.
    \item \texttt{RUMC $ViT_{256}$} $\to$ pretraining $ViT_{256}$ from random weight initialization.
\end{enumerate}

For $ViT_{4096}$, we used 343,825 $[4096\times4096]$ regions extracted from the same dataset. Starting with the extracted $[CLS]_{256}$ tokens from RUMC and TCGA+RUMC $ViT_{256}$, we conducted two separate experiments for each configuration:
\begin{enumerate}
    \item \texttt{TCGA+RUMC} $ViT_{4096}$ $\to$ finetuning $ViT_{4096}$ from the available \texttt{TCGA} $ViT_{4096}$ weights.
    \item \texttt{RUMC} $ViT_{4096}$ $\to$ pretraining $ViT_{4096}$ from random weight initialization.  
\end{enumerate}
This resulted in four distinct pretraining scenarios for $ViT_{4096}$.

\vspace{-0.3cm}
\subsection{Slide-Level Weak Supervision}
\vspace{-0.1cm}
To have an evaluation baseline, induction was run on TCGA $ViT_{256}$ and TCGA $ViT_{4096}$ to obtain the $[CLS]_{4096}$ token representations for each macro slide of the train, validation, and test sets.
Successively, we trained the last level of HIPT, $ViT_{WSI}$, to address the two classification tasks.
For binary classification, we employed cross-entropy loss during training and evaluated accuracy, precision, recall, F1 score, and the area under the curve of the receiver operating characteristics (AUC ROC), with the results presented in \cref{tab:binary_classification}.
In the multiclass classification task, focal loss was used for training due to its superior performance compared to cross-entropy loss observed in preliminary experiments. The evaluation metrics included top-1 accuracy, balanced accuracy, quadratic Cohen's Kappa, and AUC ROC, the results are detailed in \cref{tab:multiclass}. 

The results demonstrate the impact of pretraining strategies on model performance. Models pretrained without TCGA data consistently underperform, highlighting the importance of data diversity. Those utilizing TCGA+RUMC pretraining achieve the best results, emphasizing the effectiveness of incorporating task-specific data.

\begin{table}[ht]
\caption{Binary classification. AUC ROC: Area Under the Receiver Operating Characteristics Curve, Acc: Accuracy.}
\label{tab:binary_classification}
\resizebox{\columnwidth}{!}{%
\begin{tabular}{@{}l|l|c|c|c|c|c@{}}
\toprule
    \textbf {ViT$_{256}$ Pretraining} & 
    \textbf {ViT$_{4096}$ Pretraining} & 
    \textbf{Acc} & 
    \textbf{Precision} & 
    \textbf{Recall}  & 
    \textbf{F1-score} & 
    \textbf{AUC ROC} \\ \hline
TCGA & TCGA  & 0.844  & \textbf{0.988}   & 0.796 & 0.882  & 0.931 \\
TCGA+RUMC & TCGA+RUMC & 0.831 & 0.966 & 0.796 & 0.873 & \textbf{0.953} \\
TCGA+RUMC & RUMC  & \textbf{0.905}  & 0.943   & \textbf{0.925} & \textbf{0.934}  & 0.944 \\
RUMC & TCGA+RUMC  & 0.722  & 0.924   & 0.675 & 0.780  & 0.910 \\
RUMC & RUMC  & 0.614  & 0.947   & 0.500 & 0.654  & 0.841 \\ \hline
\end{tabular}%
}
\end{table}

\begin{table}[t]
\caption{Multiclass Classification. AUC ROC: Area Under the Receiver Operating Characteristics curve (we report Macro AUC), BMA: Balanced Multiclass Accuracy, Acc: Accuracy.}
\label{tab:multiclass}
\resizebox{\columnwidth}{!}{%
\begin{tabular}{@{}l|l|c|c|c|c@{}}
\toprule
  \textbf {ViT$_{256}$ Pretraining} &
  \textbf {ViT$_{4096}$ Pretraining} &
  \textbf{Acc} &
  \textbf{BMA} &
  \textbf{Kappa} &
  \textbf{AUC ROC} \\ \hline
TCGA & TCGA & 0.695 & \textbf{0.689} & 0.711 & 0.874 \\
TCGA+RUMC & TCGA+RUMC & \textbf{0.702} & 0.666 & \textbf{0.732} & \textbf{0.897} \\
TCGA+RUMC & RUMC & 0.675 & 0.618 & 0.628 & 0.858 \\
RUMC & TCGA+RUMC & 0.506 & 0.495 & 0.397 & 0.783 \\
RUMC & RUMC & 0.520 & 0.565 & 0.495 & 0.800 \\ \hline
\end{tabular}%
}
\end{table}

\vspace{-0.3cm}
\subsection{Feature representation analysis}
\begin{figure}[ht]
    \centering
    \includegraphics[width=1.0 \linewidth]{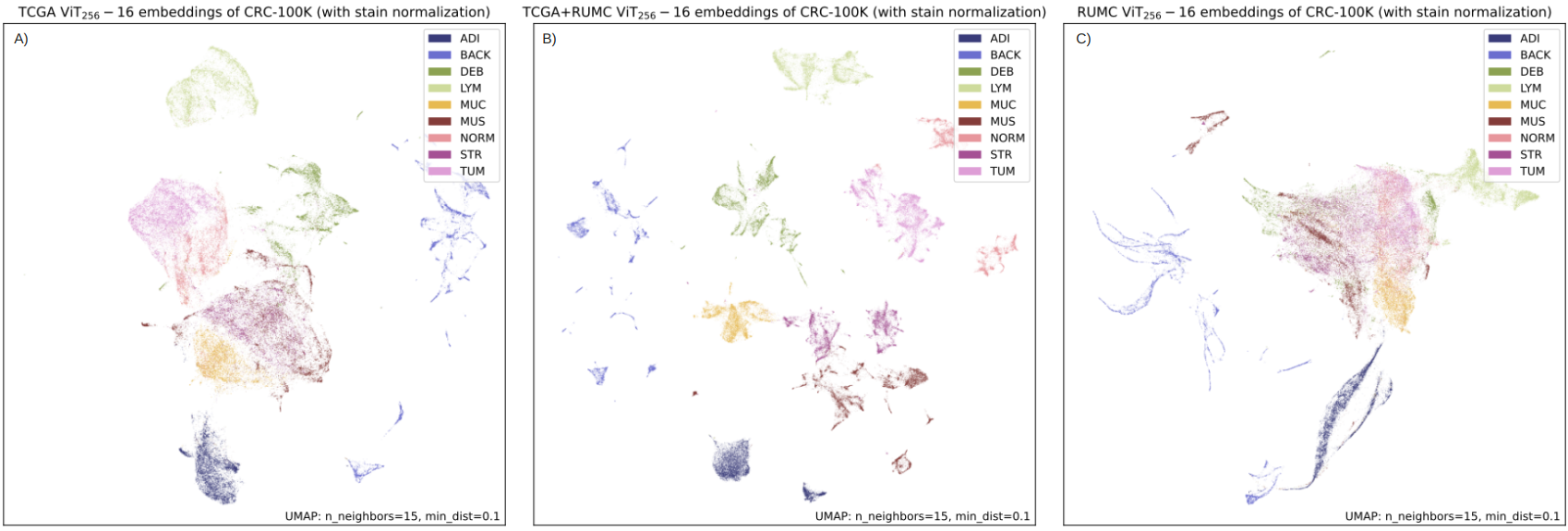} 
  \caption{UMAP plots of embeddings from A) TCGA $ViT_{256}$, B) TCGA+RUMC $ViT_{256}$ and C) RUMC $ViT_{256}$. ADI: adipose tissue, BACK: background, DEB: debris, LYM: lymphocytes, MUC: mucus, MUS: smooth muscle, NORM: normal colon mucosa, STR: cancer-associated stroma, TUM: colorectal tumor.}
  \label{fig:umaps}
\end{figure}

\vspace{-0.1cm}
We investigated the difference in representations learned by the RUMC $ViT_{256}$ encoder and the TCGA+RUMC $ViT_{256}$ encoder to the original TCGA $ViT_{256}$ through a qualitative evaluation of the features learned at this stage of HIPT.
Given that the colorectal WSI datasets used have only slide-level labels, we extracted patch-level features from the CRC-100K dataset \cite{kather_crc_dataset}. This publicly available dataset comprises 100,000 non-overlapping  $[224\times224]$ patches at 20X magnification with H\&E staining of human colorectal cancer (CRC) and normal tissue, each labeled with one of nine different tissue classes.
In \cref{fig:umaps}.a), we present the uniform manifold approximation and projection (UMAP) of features extracted from the TCGA $ViT_{256}$, revealing an almost complete overlap in the feature projections of cancer-associated stroma and smooth muscle, while the remaining class clusters are concentrated close to each other. In contrast, \cref{fig:umaps}.b) displays the UMAP of features extracted from TCGA+RUMC $ViT_{256}$, showing more compact clusters and a more even distribution of clusters across the UMAP space. This suggests that specific aspects of the data diversity in colon WSIs were effectively learned, as supported by the findings in the previous section.
Finally, \cref{fig:umaps}.c) showcases the UMAP of features from the RUMC $ViT_{256}$ encoder, where the most significant overlap between tissue clusters is observed. This overlap could potentially be attributed to the limited tissue type diversity and semantic unbalance within the RUMC cohort, primarily comprising benign cases. Due to this aspect of the employed dataset, the capability of RUMC $ViT_{256}$ to learn meaningful distinctive features without the TCGA prior appears to be weakened.

% Figure: UMAP

\vspace{-0.3cm}
\section{Conclusion}
\vspace{-0.1cm}
We have presented an exploration of self-supervised knowledge in HIPT to learn representations of colon biopsy in WSIs and address classification tasks in this domain.
We have shown the superiority of a TCGA-pretrained model fine-tuned on domain-specific data (i.e., TCGA+RUMC) for classification tasks, and reported inferior performance of a model trained from scratch with domain-specific data only (i.e., RUMC data). 
The RUMC dataset originates from screening studies, thereby encompassing a significant proportion of benign (normal) cases. In contrast, the TCGA dataset is characterized by the inclusion of representative cancer types. Given that our downstream tasks are on cancer classification, it is evident that pretraining on a large cohort of different cancer types imbues HIPT with robust cancer-discriminative capabilities. However, it remains uncertain how TCGA pretraining may perform on a non-cancer related task (i.e. celiac disease). This highlights the importance of variety in the data used for self-supervised pretraining, crucial for its good performance in different downstream tasks. Future work should focus on developing sampling techniques to have a balanced representation of the different tissue types contained in datasets.

\textbf{Compliance with ethical standards:}
All procedures performed in studies involving human participants were in accordance with the ethical standards of the institutional and/or national research committee.

\textbf{Acknowledgments:} 
This project was partly funded by the European Union's Horizon 2020 research and innovation programme under grant agreement No 825292 (ExaMode, htttp://www.examode.eu/). NSLC would like to express her appreciation for the \emph{Erasmus Mundus Joint Master Degree in Medical Imaging and Applications} and the \emph{Radboud AI for Health} projects for their financial support and their commitment to promoting artificial intelligence research that solves clinical problems.
FC was Chair of the Scientific and Medical Advisory Board of TRIBVN Healthcare, France, and received advisory board fees from TRIBVN Healthcare, France in the last five years. He is shareholder of Aiosyn BV, the Netherlands. All other authors declare no conflict of interest.

\bibliographystyle{IEEEbib}
\vspace{-0.3cm}
\bibliography{refs}

\end{document}